\documentclass[12pt]{iopart}
\usepackage{iopams}  

\usepackage{graphics}
\usepackage[sort]{cite}

\begin{document}
\title{Illustration of quantum complementarity using single photons interfering on a grating}
\author{V~Jacques$^{1}$, N~D~Lai$^{1}$, A~Dr\'eau$^{1}$, D~Zheng$^{1}$, D~Chauvat$^{1}$, F~Treussart$^{1}$, P~Grangier$^{2}$, and J-F~Roch$^{1}$}

\address{$^{1}$
Laboratoire de Photonique Quantique et Mol\'eculaire, Ecole Normale Sup\'erieure de Cachan, UMR CNRS 8537, Cachan, France}

\address{$^{2}$
Laboratoire Charles Fabry de l'Institut d'Optique, UMR CNRS 8501, Palaiseau, France}

\ead{roch@physique.ens-cachan.fr}

\begin{abstract}
A recent experiment performed by S. S. Afshar {\it et al.} has been interpreted 
as a violation of Bohr's complementarity principle between interference visibility 
and which-path information in a two-path interferometer. We have reproduced this 
experiment, using true single-photon pulses propagating in a two-path wavefront- 
splitting interferometer realized with a Fresnel's biprism, and followed by a grating 
with adjustable transmitting slits. The measured values of interference visibility $V$ 
and which-path information, characterized by the distinguishability parameter $D$, are found to obey the complementarity relation $V^{2}+D^{2}\leq 1$. This result demonstrates that the experiment can be perfectly explained by the Copenhagen interpretation of quantum mechanics.
\end{abstract}

 % end of PACS codes

\submitto{\NJP}

\maketitle
\section{Introduction}

Bohr's principle of complementarity states that every quantum system has mutually 
incompatible properties which cannot be simultaneously measured~\cite{Bohr}. This principle is commonly illustrated by considering single particles in a two-way interferometer where one chooses either to observe interference fringes, associated to a wave-like behavior, or to know which path of the interferometer has been followed, according to a particle-like behavior~\cite{Feynman}. In such an experiment, any attempt to obtain some which-path information (WPI) unavoidably reduces interference, and reciprocally. The incompatibility between these two measurements is then ensured by the complementarity inequality~\cite{Jaeger_PRA1995,Englert}: 
\begin{equation}
V^{2}+D^{2}\leq 1  
\label{ComplIneq}
\end{equation}
which puts an upper bound to the maximum values of independently determined 
interference visibility $V$ and path distinguishability $D$, the parameter that quantifies the available WPI on the quantum system~\cite{Englert}.

\indent The two all-or-nothing cases $(V=1,D=0)$ and $(V=0,D=1)$ have been clearly confirmed by experiments performed with a wide range of quantum objects~\cite{Tonomura,Neutrons,Carnal,Keith,Arndt,Pfau,Pritchard,RempeNature,Buks,Grangier,Jacques}, as well in the quantum eraser configuration~\cite{Brune_PRL1997,Scully,Haroche} or in Wheeler's delayed-choice regime~\cite{DCECachan}. The complementarity inequality~(\ref{ComplIneq}) has also been successfully verified in intermediate regime, corresponding to partial WPI and reduced visibility, with atoms~\cite{Haroche,Rempe}, nuclear spins~\cite{Peng}, and single photons in the delayed-choice regime~\cite{DCEComp}. Although recent discussions focused on the mechanism which enforces complementarity, by discussing its relation with Heisenberg's uncertainty relations~\cite{ScullyNature,EnglertBis,WisemanBis,Wiseman}, it is well established that Bohr's complementarity principle is a cornerstone of quantum mechanics~\cite{Wheeler}.

\indent Recently, Afshar {\it et al.} have claimed to be able to violate this principle~\cite{Afshar1,Afshar2}. Their experimental scheme, depicted on figure 1, can be summarized as follows: Attenuated laser light illuminates a Young's double-pinhole screen which produces an interference pattern at a distance behind the two pinholes $S_{1}$ and $S_{2}$ where the two diffracted beams overlap. Using a lens, each pinhole is imaged on an associated detector, {\it i.e.} $S_{1}$ on $\rm P_{1}$ and $S_{2}$ on $\rm P_{2}$. Each detector is then univocally associated to a given path of the interferometer, leading to the full knowledge of the WPI  and corresponding to $D=1$.

\indent In order to simultaneously recover the complementary wave-like information, a grid of thin wires is inserted close to the imaging lens. The wires are exactly superposed 
on the dark fringes of the interference pattern (see figure 1). Using a particle-like 
description, Afshar {\it et al.} claim that no photon is blocked by the grid and the 
signals associated to the output detectors $\rm P_{1}$ and $\rm P_{2}$ remain almost unchanged, as experimentally verified. Their conclusion is that the grid of wires perfectly reveals the interference pattern while keeping a perfect WPI, corresponding to combined 
measurements of $V=1$ and $D=1$. This result, in clear contradiction with inequality~(\ref{ComplIneq}), is interpreted as a violation of the complementarity principle.

\indent Different papers have pointed out the flaws in the interpretation of the experiment and explained why there is no contradiction with Bohr's complementarity~\cite{AfsharRefut}. In this paper, we report an experiment designed to check the complementarity inequality using a setup similar to the one of figure 1, the Young's double-pinhole being replaced by a Fresnel's biprism. To be meaningful, the experiment is realized with true single-photon pulses for which full and unambiguous WPI can be obtained, complementary to the observation 
of interference~\cite{SPS}. 

\indent The paper is organized as follows: We start with a wave-like analysis of the experiment, allowing us to determine the interference visibility $V$  and the path distinguishability parameter $D$. We demonstrate that the set of these two parameters obeys inequality~(\ref{ComplIneq}). This analysis is then compared to the experiment. The results correspond to the almost ideal case, close  to the upper bound of inequality~(\ref{ComplIneq}).

%%%%%%%%%% figure1 %%%%%%%%%%%%
\begin{figure}[ta]
\centerline{\resizebox{0.6\columnwidth}{!}{
\includegraphics{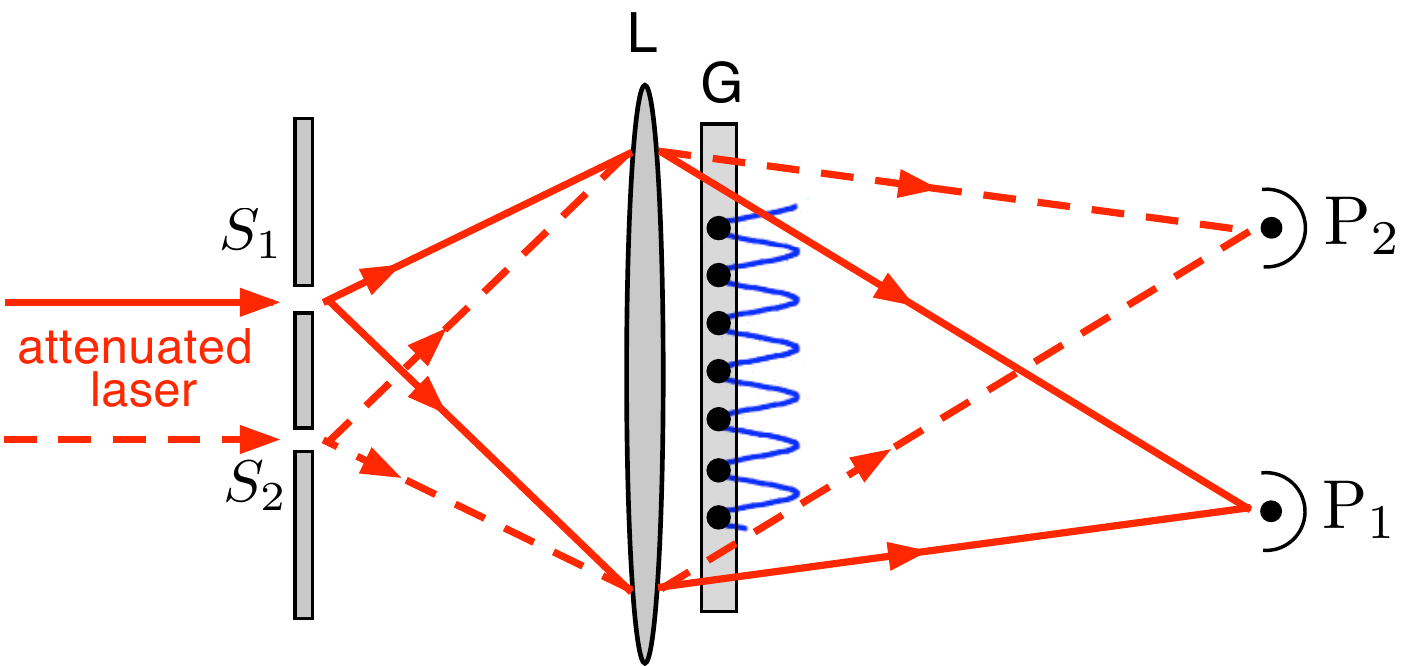}}}
\caption{Simplified representation of Afshar's experiment~\cite{Afshar1,Afshar2}. An attenuated laser illuminates a Young's double-pinhole interferometer. A lens (L) images each pinhole $S_{1}$ and $S_{2}$ on two detectors $\rm P_{1}$ and $\rm P_{2}$. A grid of thin wires (G) with a period matching the interfringe is positioned after the lens so that the wires of the grid are exactly superimposed on the dark fringes of the interference pattern. }
\end{figure}
%%%%%%%%%%%%%%%%%%%%%%%%

\section{Afshar's setup with a Fresnel's biprism: A wave-optical analysis}
\label{WOA}

\indent Figure 2-(a) shows the setup corresponding to two separated incident beams at normal incidence on a Fresnel's biprism, with two output detectors $\rm P_{1}$ and $\rm P_{2}$ positioned far away from the overlapping region of the two deviated beams~\cite{Jacques}. Each detector is then unambiguously associated to a given path of the interferometer, {\it i.e.} detector $\rm P_{1}$ to path 1 and detector $\rm P_{2}$ to path 2. The experiment depicted on figure 1 can then be reproduced by introducing a transmission grating inside the interference zone corresponding to the overlap of the two beams refracted by the biprism.

\indent A strong assumption in Afshar's interpretation is that positioning the wires of the grid at the dark-fringe locations is enough to reveal the existence of the interference pattern, without inducing any further perturbation on the transmitted light field. However, the grid has an unavoidable effect due to diffraction, which redirects some light from path 1 to detector $\rm P_{2}$ and, reciprocally, from path 2 to detector $\rm P_{1}$. The introduction of the grid has then partially erased the WPI since it becomes  impossible to univocally associate each output detector to a given path of the interferometer. 

\indent We first need to evaluate the influence of diffraction due to the grating G. As shown in figure 2-(b), G corresponds to transmitting slits of width $a$ with a periodicity equal to the interfringe $\Lambda$ of the interference pattern obtained with monochromatic light of wavelength $\lambda$. The interfringe depends on the deviation angle $\alpha_{0}=(n-1)\beta$ caused by the Fresnel's biprism of refraction index $n$ and summit angle $\beta$:\begin{equation}
\Lambda=\frac{\lambda}{2\alpha_{0}}=\frac{1}{2u_{0}} \ ,
\label{InterfBiprisme}
\end{equation}
\noindent when expressed as a function of the associated spatial frequency $u_{0}=\alpha_{0}/\lambda$.

\begin{figure}[t]
\centerline{\resizebox{0.8\columnwidth}{!}{
\includegraphics{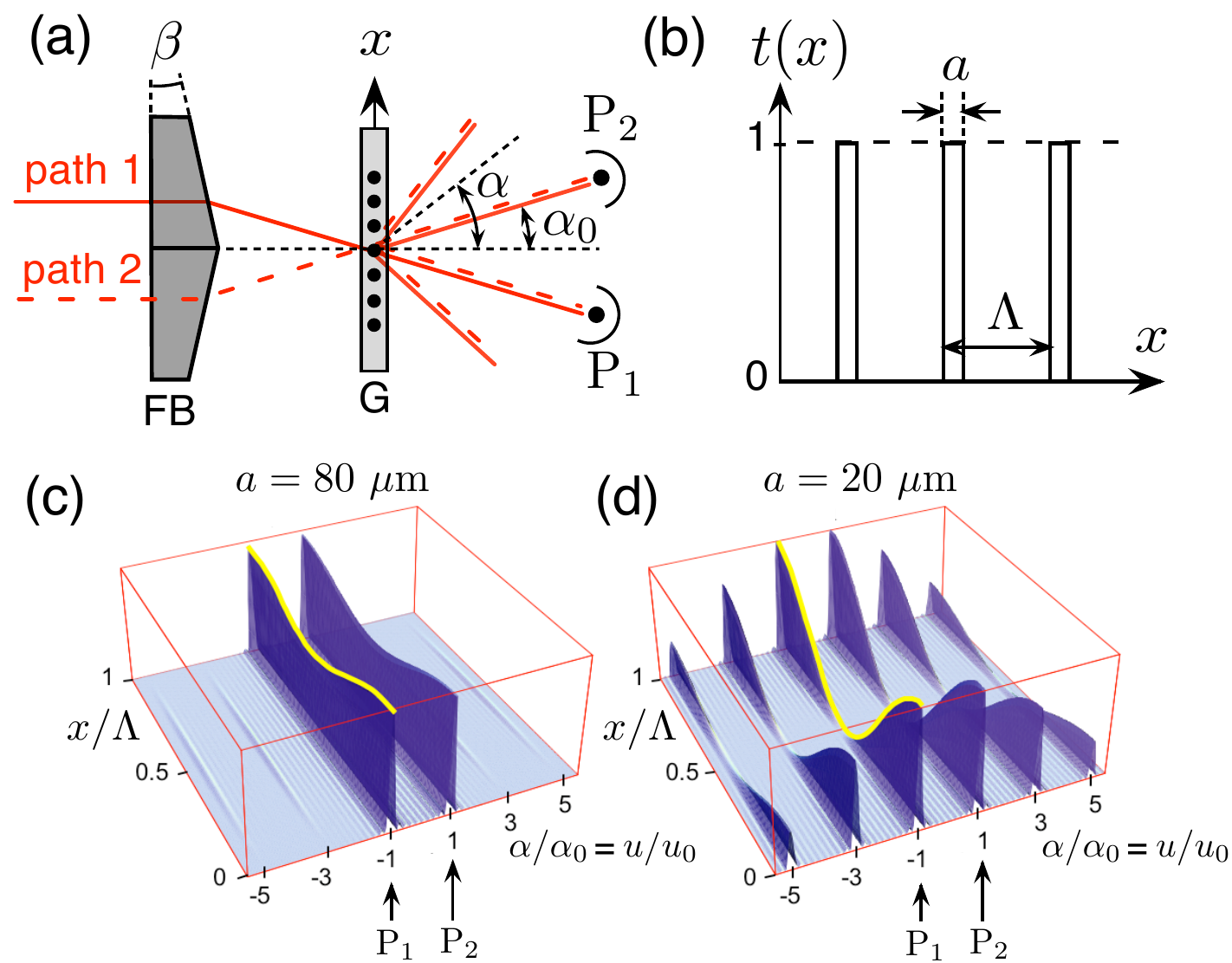}}}
\caption{Modified Afshar's experiment with a Fresnel's biprism (FB) of summit angle $\beta$ and two interfering paths 1 and 2. (a)-Two detectors $\rm P_{1}$ and $\rm P_{2}$ are positioned far away from the interference area and are therefore each univocally associated to a given path of the interferometer. A grating (G) is then introduced in the interference area and can be moved along the $x$-axis of the interference pattern. (b)-G is modelized as an amplitude transmission function $t(x)$ with periodicity $\Lambda$ and transmitting slits of width $a$. (c)-(d) Light intensity distribution after diffraction by the grating G as a function of angle $\alpha$ and grating position $x$, for transmitting slit width values $a=80 (c)\ \mu$m and $20 \ \mu$m (d). Light intensities of all diffraction orders undergo maxima and minima when G is translated from a bright interference fringe $(x = p\Lambda, \ p = 0, 1, 2\dots)$ to a dark interference fringe $(x = p\Lambda+\Lambda/2, \ p =1, 2\dots)$. The detectors $\rm P_{1}$ and $\rm P_{2}$ are respectively associated with propagation at oblique angle $\alpha=-\alpha_{0}$ ($u=-u_{0}$) and $\alpha=\alpha_{0}$ ($u=u_{0}$)( black arrows). The calculation is done with $\beta=7.5\times 10^{-3}$ rad, $\Lambda= 87 \ \mu$m, and $N=20$, corresponding to the values of the experiment described in the latter.}
\label{exp_setup}
\end{figure}

\indent As well known from quantum optics~\cite{Gerry}, all optical phenomena like interference, diffraction, and propagation, can be calculated using the classical theory of light even in the single-photon regime. Then, using classical-wave Fraunhofer diffraction, the diffracted wave amplitudes $\mathcal{S}_{1}(u)$ and $\mathcal{S}_{2}(u)$ associated to path 1 and path 2 of the interferometer are:
\begin{equation}
\mathcal{S}_{1}(u) =\mathcal{S}_{0} \ {\rm sinc} [\pi(u+u_{0})a]  \frac{\sin[N\pi(u+u_{0})\Lambda]}{\sin[\pi(u+u_{0})\Lambda]} {\rm e}^{i\pi(N-1)\left[\frac{u-u_{0}}{2u_{0}}\right]}  {\rm e}^{-2i\pi(u-u_{0})x}
\label{S1}
\end{equation}
\begin{equation}
\mathcal{S}_{2}(u) =\mathcal{S}_{0} \ {\rm sinc} [\pi(u-u_{0})a]  \frac{\sin[N\pi(u-u_{0})\Lambda]}{\sin[\pi(u-u_{0})\Lambda]} {\rm e}^{i\pi(N-1)\left[\frac{u+u_{0}}{2u_{0}}\right]}   {\rm e}^{-2i\pi(u+u_{0})x}
\label{S2}
\end{equation}
\noindent where $u=\alpha/\lambda$ is the spatial frequency associated to propagation with oblique angle $\alpha$, $x$ is the position of the grating along the $x$-axis and $N$ is the number of transmitting apertures illuminated by the incident beams of equal amplitude $\mathcal{S}_{0}$.

\indent Consequently, detector $\rm P_{1}$ (resp. $\rm P_{2}$) positioned in direction $u=-u_{0}$ (resp. at $u=u_{0}$) is associated to the zero-order diffraction (resp. first-order)  from path 1 and also to the first-order diffraction (resp. zero-order) from path 2. The WPI on the behavior of a single-photon in the interferometer is then partially erased as each detector cannot be associated to a given path.\\
\indent To test inequality~(\ref{ComplIneq}), a value of the distinguishability parameter $D$ is required, to quantify the amount of WPI that can be extracted in the experiment. Following the discussion of reference~\cite{DCEComp}, we introduce the parameters $D_{1}$ and $D_{2}$, respectively associated to the WPI on path 1 and on path 2: 
\begin{equation}
D_{1}=\left|p({\rm P}_{1},{\rm path} \ 1)-p(\rm P_{2},\rm path \ 1)\right| \ ,
\end{equation}
\begin{equation}
D_{2}=\left|p({\rm P}_{1},{\rm path} \ 2)-p(\rm P_{2},\rm path \ 2)\right| \ ,
\label{defI1}
\end{equation}
\noindent where $p(\rm P_{i},\rm path \ j)$ is the probability that the particle follows path j and is detected on detector $\rm P_{i}$. The distinguishability parameter $D$ is then finally defined as~\cite{Englert}:
\begin{equation}
D=D_{1}+D_{2}.
\label{Predic}
\end{equation}

\indent Using true single-photon pulses and photodetectors operating in the photon counting regime, the values of $D_{1}$ and $D_{2}$ can be estimated by blocking one path of the interferometer and measuring the corresponding number of detections $N_{1}$ and $N_{2}$ on detectors $\rm P_{1}$ and $\rm P_{2}$. This quantities are statistically related to $D_{1}$ and $D_{2}$ according to~\cite{Rempe,DCEComp} :
\begin{equation}
D_{1}=\frac{1}{2}\left.\left|\frac{N_{1}-N_{2}}{N_{1}+N_{2}} \right| \right]_{\rm path \ 2 \  blocked}  \, ,
\label{D1exp}
\end{equation}
\begin{equation}
D_{2}=\frac{1}{2}\left.\left|\frac{N_{1}-N_{2}}{N_{1}+N_{2}} \right| \right]_{\rm path \ 1 \  blocked} \, .
\label{D2exp}
\end{equation}
Using equations~(\ref{S1}) and~(\ref{S2}), the distinguishability parameter $D$ is then equal to:
 \begin{equation}
D= \frac{1-{\rm sinc}^{2}(2\pi u_{0}a)}{1+{\rm sinc}^{2}(2\pi u_{0}a)} \ .
\label{D}
\end{equation}

\indent In the extreme case of a grating consisting of Dirac transmission peaks (equivalent to the limit case $a=0$), $D$ is equal to zero and no WPI can be obtained. Conversely, when the grating is absent, equivalent to the case where $a=\Lambda$, we obtain $D=1$.\\
\indent In order to retrieve the complementary wave-like information associated to the single-photon detections on detectors $\rm P_{1}$ and $\rm P_{2}$, a quantitative measurement of the interference visibility is required. Note that such measurement cannot be realized by positionning the grating as described in references~\cite{Afshar1,Afshar2}. Indeed, the visibility inferred using such method is related to photons intercepted by the grid for which no WPI is available. We stress that the complementarity inequality is only meaningful if both complementary measurements are performed for the {\it same} photons, {\it i.e.} either for photons transmitted behind the grating or for photons intercepted by the grating.  \\
\indent Using detectors $\rm P_{1}$ and $\rm P_{2}$, the wave-like information complementary to the WPI defined by equation~(\ref{D}) can be measured by translating the grating along the $x$-direction (see figure 2-(a)). The intensity $\mathcal{I}(u)$ of diffracted light behind the grating is given by:
\begin{eqnarray}
\mathcal{I}(u)&= \left| \mathcal{S}_{1}(u)+\mathcal{S}_{2}(u) \right|^{2}\\
 &= \left| \mathcal{S}_{1}(u) \right|^{2}+\left| \mathcal{S}_{2}(u) \right|^{2}+2 {S}_{1}(u) {S}_{2}^{\ast}(u) \cos(4\pi u_{0}x)
 \label{Vth}
\end{eqnarray}
The counting rates on detectors $\rm P_{1}$ and $\rm P_{2}$ are then modulated as a function of the grating position $x$ (see figure 2-(c)), corresponding to an interference visibility:

 \begin{equation}
V= \frac{2{\rm sinc}(2\pi u_{0}a)}{1+{\rm sinc}^{2}(2\pi u_{0}a)}
\label{V}
\end{equation}

\indent In the limit $a=0$, the visibility is equal to unity whereas it becomes null in the absence of grating ($a=\Lambda$).

\indent Combining equations~(\ref{D}) and~(\ref{V}) leads to $V^{2}+D^{2}=1$, in agreement with inequality~(\ref{ComplIneq}). Opposite to the conclusion that it leads to a violation of Bohr's complementarity, the experimental setup proposed in references~\cite{Afshar1,Afshar2} provides a nice and clever illustration of the balance between which-path information and interference visibility in a two-path interferometer. Note that this experiment differs from usually considered which-way schemes consisting of an interferometer where one tries to get (either a priori or a posteriori) information about the path followed by the particle, leading to a degradation of  the interference visibility according to inequality~(\ref{ComplIneq}). In our experiment, we start from a perfect which-way knowledge and we try to conversely retrieve the wave-like information. As it can be expected, this slight change in perspective does not affect the complementarity argument, and the usual reasoning made above with the quantities $V$ and $D$ remains therefore valid.

\section{Experimental results}

\indent The above predictions are now compared with the experiment which setup is shown in figure 3. The experiment starts from a clock-triggered single-photon source based on the photoluminescence of a single NV colour centre in a diamond nanocrystal~\cite{Kun,DCECachan}. Since the photoluminescence spectrum of NV colour centre is very broad (about $100$-nm FWHM at room temperature), we use a $10$-nm-bandwidth bandpass filter centered at $\lambda=670$ nm, corresponding to the emission peak of the NV centre. This spectral filtering allows us to extend the coherence length of the single-photon pulses. The linearly-polarized single-photon pulses are divided into two spatially separated paths of equal amplitudes, using polarization beam displacers (BD) and half-wave plates. The experimental configuration leads to a $5.6$-mm beam separation while keeping zero optical path difference between the two interfering channels. A third half wave plate is then selectively placed in one beam, in order to obtain two beams with identical polarizations (see figure 3). The optical path difference induced by this half-wave plate is compensated by a piece of glass (GP) introduced on the other beam. After preparation, the two single-photon beams have then identical linear polarization state, equal amplitude, no optical path difference between them and a spatial separation high enough to avoid any diffraction effect at the apex of the Fresnel's biprism. The prepared single-photon beams are finally sent at normal incidence through the Fresnel's biprism followed by the transmission grating. The diameter of each beam is around $2$~mm, corresponding to the illumination of approximately $20$ slits of the grating.

\begin{figure}[t]
\centerline{\resizebox{0.9\columnwidth}{!}{
\includegraphics{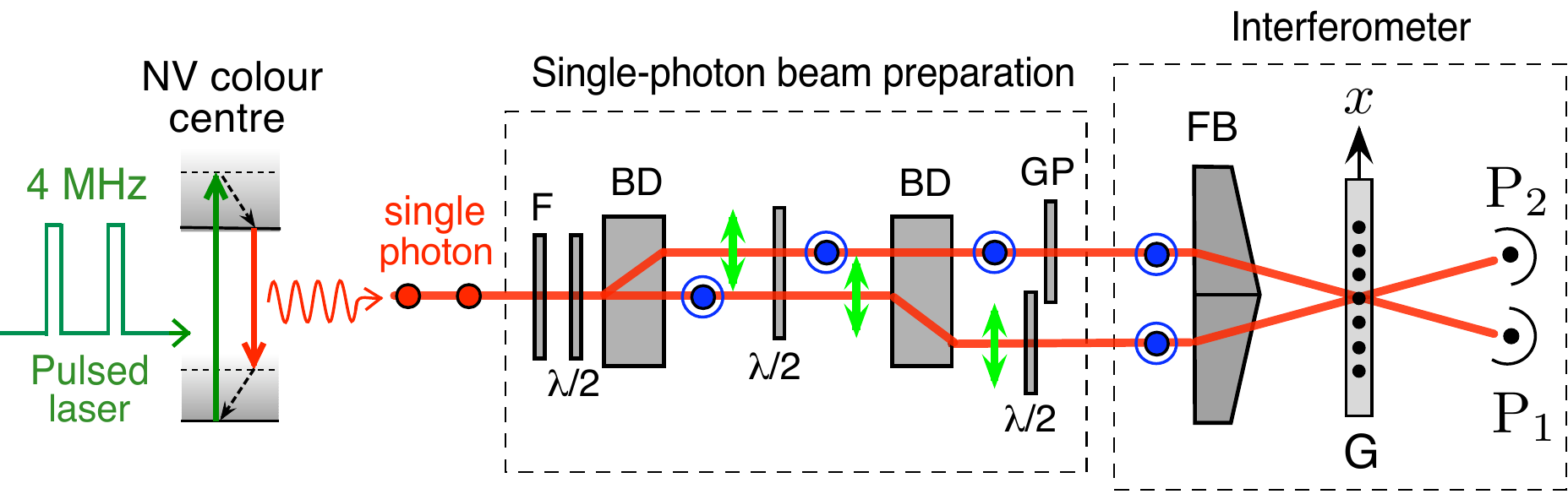}}}
\caption{Experimental realization of the modified Afshar's experiment based on a Fresnel's biprism and single-photon pulses emitted by an individual NV colour centre in a diamond nanocrystal, excited in pulsed regime at a $4$-MHz repetition rate. $\lambda /2$: half-wave plate. BD: YVO$_{4}$ polarized beam displacer. GP: glass plate. F: $10$-nm-bandwidth bandpass filter centered at $\lambda=670$ nm. FB: Fresnel's biprism. G: transmission grating inserted in the interference area and translated along the interference $x$-axis. $\rm P_{1}$ and $\rm P_{2}$: avalanche photodiodes positionned in the zero-order diffraction direction of each beam and operated in the photon-counting regime (Perkin Elmer, AQR14). }
\end{figure}

\indent As meaningful illustration of complementarity requires the use of single particles~\cite{Grangier}, the quantum behavior of the light field is first tested using the two output detectors feeding single and coincidence counters without the grating. In this situation, we measure the correlation parameter $\alpha$~\cite{Grangier,Jacques} which is equivalent to the second-order correlation function at zero delay $g^{(2)}(0)$. For an ideal single-photon source, quantum optics predicts a perfect anticorrelation $\alpha=0$, in agreement with the particle-like image that the photon cannot be detected simultaneously in the two paths of the interferometer. With our source, we find $\alpha = 0.14 \pm 0.02$. This value, much smaller than unity, shows that we are close to the pure single-photon regime~\cite{alpha}.

\begin{figure}[b]
\centerline{\resizebox{0.8\columnwidth}{!}{
\includegraphics{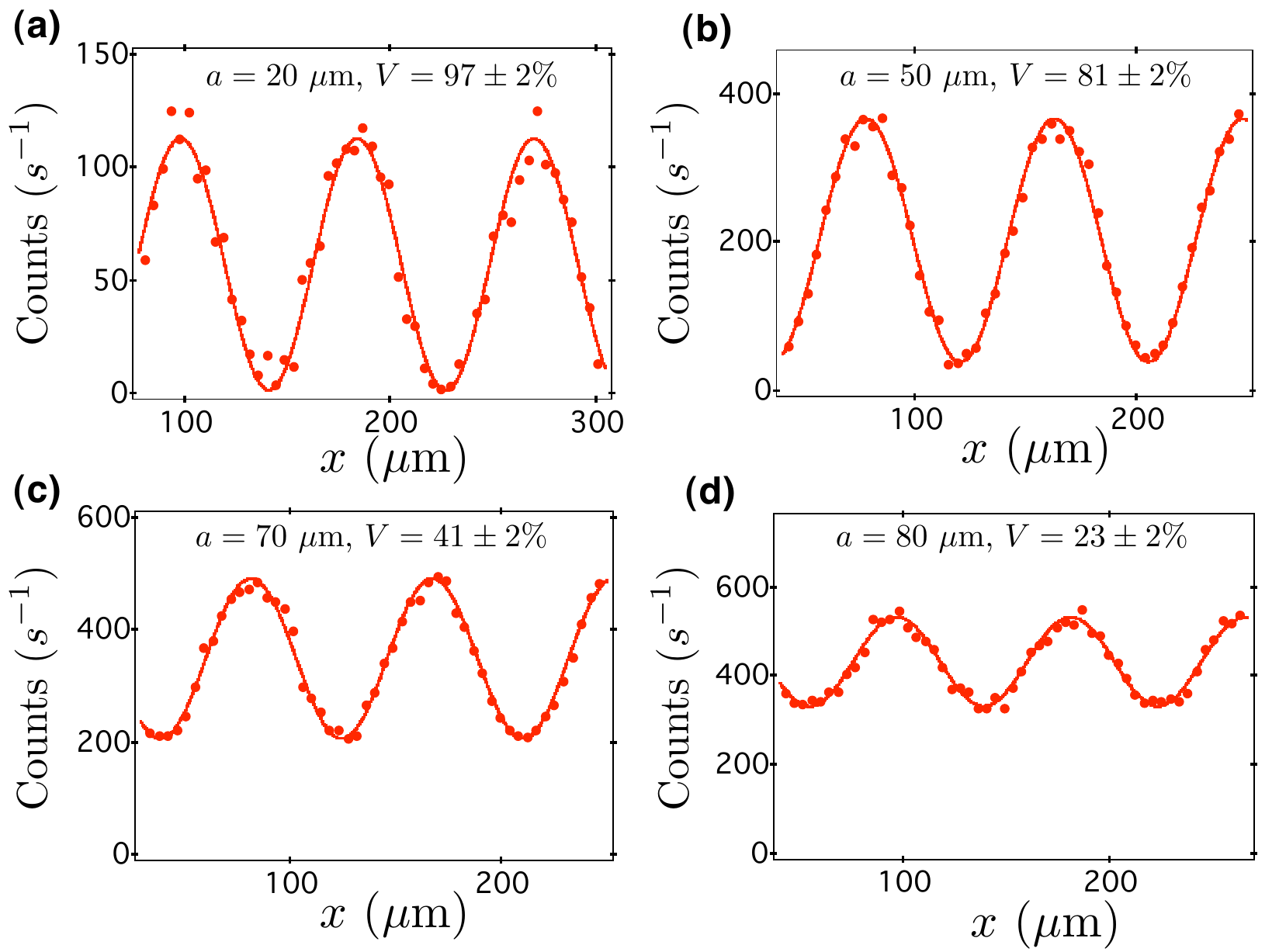}}}
\caption{Photocounts recorded on detector $\rm P_{1}$ while translating the grating G along the $x$-axis, for different widths $a$ of the transmitting slits: (a) $a = 20 \ \mu$m, (b) $ 50 \ \mu$m, (c) $ 70 \ \mu$m and (d) $ 80 \ \mu$m. Identical result are recorded on detector $\rm P_{2}$. The grating is translated by   $4$-$\mu\rm m$ steps and each point is recorded with $3$-s acquisition time. A constant averaged background due to detector dark count rate (about $180 \ \rm counts.s^{-1}$) has been subtracted from the data. The visibility is evaluated using a fit by a cosine function shown in solid line.}
\end{figure}

\indent With the parameters of the Fresnel's biprism ($\beta=7.5\times 10^{-3}$ rad and $n=1.51$), the interfringe is $\Lambda=87 \ \mu$m at $\lambda=670$ nm (see equation~(\ref{InterfBiprisme})). Following the discussion of section~\ref{WOA}, we then use a set of gratings with the same period $\Lambda$ but with different widths $a$ of the transmitting slits. The experiment then consists in measuring $D$ and $V$ for each value of the parameter $a$.

\indent Each grating is introduced into the interference area and translated along the $x$-axis of the interference pattern, using a computer-driven translation stage with sub-micrometer accuracy positioning.
 As shown in figure 4, a modulation of the counting rates is observed for detectors $\rm P_{1}$ and $\rm P_{2}$, allowing us to estimate the wave-like information by measuring the visibility of the modulation for each grating. As expected, the visibility $V$ decreases when the width $a$ of the transmitting slits increases, the dependance being in good agreement with equation~(\ref{D}) (see figure 5-(a)). \\
 \indent For each grating, we independently measure the distinguishability parameter $D$ to quantify the available WPI. This is experimentally realized by consecutively blocking one arm of the interferometer and the other, and by measuring the quantity $D_{1}$ and $D_{2}$ defined by equations~(\ref{D1exp}) and~(\ref{D2exp}). The final results, shown on figure 5-(a), lead to $V^{2}+D^{2}=0.96 \pm 0.03$ (see figure 5-(b)), close to the maximal value permitted by inequality~(\ref{ComplIneq})  even though each quantity varies from zero to unity.
 
\indent Note that when the width $a$ of the transmitting slits is wide ($a = 80 \ \mu$m), which corresponds to the setup of references~\cite{Afshar1,Afshar2} with very thin wires, the wave-like information associated to single-photon detection on detectors $\rm P_{1}$ and $\rm P_{2}$ is equal to $V^{2}=0.05\pm0.01$, very far from unity. This result illustrates that a quantitative measurement of the visibility of the transmitted light is necessary to qualify the wave-like information in a two-path interference experiment, which the simple positioning of the grid of wires at the dark fringes of the interference pattern does not realize.

\begin{figure}
\centerline{\resizebox{0.75\columnwidth}{!}{
\includegraphics{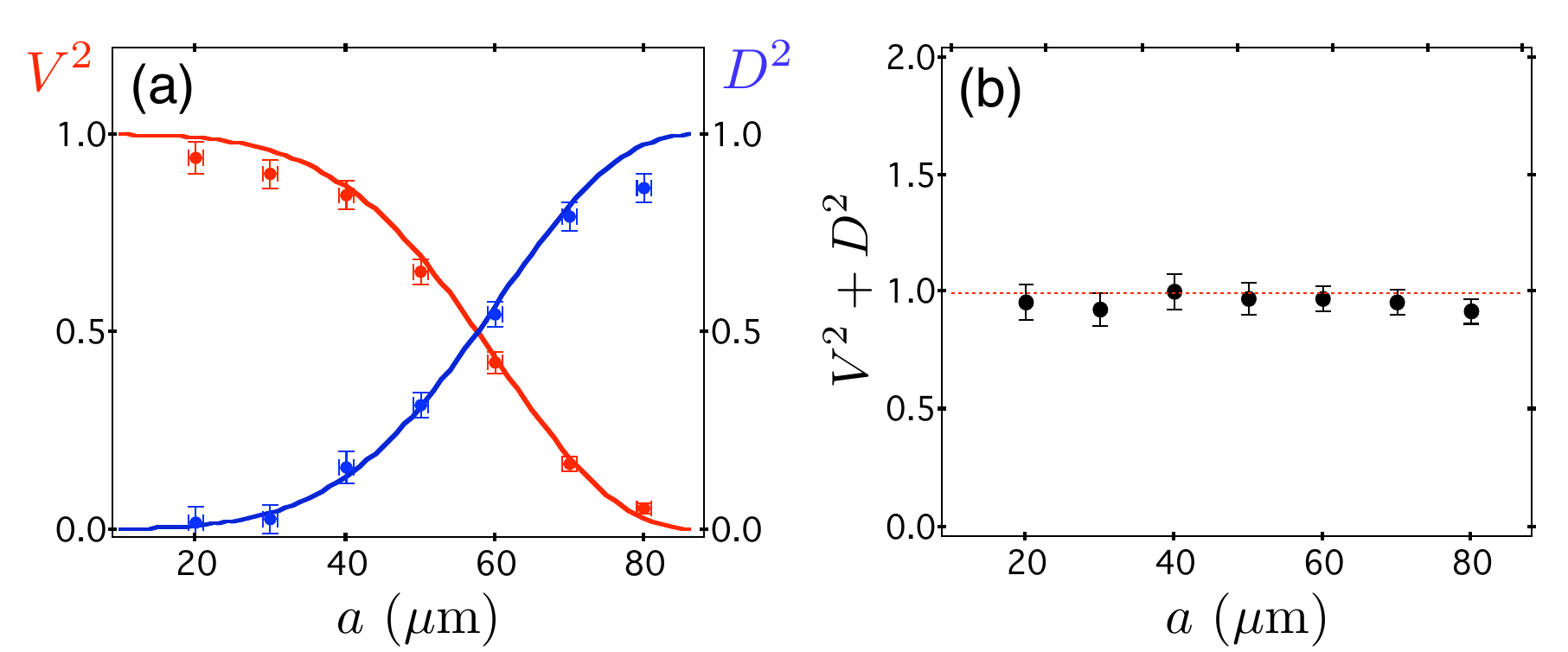}}}
\caption{(a) Wave-like information $V^{2}$ and particle-like information $D^{2}$ as a function of the width $a$ of the transmitting slits. The solid lines are the theoretical expectations given by equations~(\ref{D}) and~(\ref{V}) without any fitting parameter. (b) $D^{2}+V^{2}$ as a function of $a$.}
\end{figure}

\section{Conclusion}

We have reported the two complementary measurements of ``interference   vs. which-path information'' using single-photon pulses and a setup close to the proposal by Afshar {\it et al}~\cite{Afshar1,Afshar2}. By investigating intermediate situations corresponding to partial path distinguishability  and reduced interference visibility, we have shown that the results are in perfect agreement with the complementarity inequality. While the results may not appear as a big surprise, we hope that our experiment can contribute to clarify confuse debates around the dual behavior of the lightfield, which in Feynman's words contains \textit{``the only mystery of quantum mechanics''}~\cite{Feynman}. So far, Bohr's complementarity principle has thus never been violated.

\ack{
The authors are grateful to A~Aspect and F~Grosshans for many fruitful discussions and to S~R\'egni\'e and F~Record for their contributions to early stages of the experiment. We thank G~Colla and R~Mercier (Institut d'Optique Graduate School) for the realization of the Fresnel's biprism, and J-P Madrange (ENS Cachan) for all mechanical realizations. This work is supported by Institut Universitaire de France and European projects EQUIND (FP6 project number IST-034368) and NEDQIT (ERANET Nano-Sci).}

\end{document}